\documentclass{article} 
\usepackage{dafne99pro} 
\begin{document} 
\title{
CHARM RESULTS ON CP VIOLATION AND MIXING } 
\author{Jeffrey A. Appel 
        \\
{\em Fermilab, Batavia, IL, 60510, USA} \\
}
\maketitle
\baselineskip=11.6pt
\begin{abstract}
The most recent results on CP violation and mixing in the charm system are
reviewed as a guide to the future.  While no surprising results are reported so
far, charm provides a unique window to physics beyond the Standard Model.  The
results reported here come from four sources: ALEPH at LEP, E791 and FOCUS/E831 at
Fermilab, and CLEO II.V at CESR.  Results beyond these sources may be expected 
as a byproduct of B-motivated experiments.
\end{abstract}
\baselineskip=14pt
\section{Introduction}

So far, there is no evidence for either CP violation or particle-antiparticle
mixing in the charm-quark sector.  This is as expected in the context of the 
Standard  Model of particle physics.  Predictions for CP violation and
particle-antiparticle mixing are orders of magnitude below the 
sensitivities of current experiments.  This remains true even though 
today's experiments are part of the march toward Standard-Model 
sensitivities, a march which has seen a couple of orders of magnitude 
increase in sensitivity in each of the last two decades.

\subsection{The Present as a Guide to the Future}
 
What would be really exciting is the observation of  CP violation or 
particle-antiparticle mixing in current charm experiments.  The Standard-Model 
predictions which explain these effects for strange and bottom quarks typically predict 
(so far) unmeasurable effects for charm.  In that case, experimental charm signatures 
have no Standard-Model background, no relevant hadronic uncertainty in background 
estimates.  Any sighting of CP violation or mixing in the charm sector would be
evidence of new physics.  Since no such sighting has been made, we must settle, 
for now, simply to use the current experimental efforts as guides to future 
possibilities.  How can we best pursue the search for CP violation
and particle-antiparticle mixing?

Today's results come from four sources of charm particles: $e^+e^-$ 
experiments in the upsilon region (CLEO II.V at CESR) and at the $Z^o$ 
(ALEPH at LEP), photoproduction (FOCUS/E831 at Fermilab), 
and hadroproduction  (E791 at Fermilab).  From the next generation of experiments, 
we may hope for a continuation of increased sensitivity - though in a more limited 
set of experimental environments.

\subsection{Charm, a Unique Window to New Physics}

While the charm-physics sector has no measurable Standard Model mixing or CP
violation, it is unique in much more interesting ways than simply having
no Standard-Model backgrounds.  It is the only opportunity to see
new-physics coupling to the up-type-quark sector.    In the case of the
up quarks themselves, there is a lack of sufficient particle lifetime and
richness in decay channels for CP violation or particle-antiparticle mixing to be
manifest.  As for the top quark, it doesn't live long enough to be included in
particles which can mix or can have the final state interactions needed to see CP
violation.

The smallness of the Standard-Model diagrams gives insight into 
the uniqueness of the charm sector.  Possible contributions from box, 
penguin, and long distance effects are usually all about same 
order.\cite{ptrov}
Even when long distance effects are thought to be larger than 
perturbative Standard-Model effects, the predictions are still many 
orders of magnitude from present limits.\cite{georgiohl}  Any of a long
list of non-Standard-Model sources could produce measurable mixing or CP
violation in charm.\cite{nelsonpred}\cite{burdman}\cite{hewett}\cite{nir} 
These include leptoquarks, SUSY particles, fourth-generation quarks, 
left-right symmetric particles, and Higgs particles.

\section{Particle-Antiparticle Mixing}

Particle-antiparticle mixing can occur only for neutral particles, such as
the  $D^o$.  Three types of measurements have been made: those 
using hadronic decays, those using semileptonic decays, and those 
where comparisons are made in the decay rates to various mixtures of
CP eigenstates.  In the first two of these, one needs to know the 
nature of the  $D$ meson when it is born, i.e., produced.  Such mesons are 
referred to as tagged (as to particle-antiparticle nature at birth).  
We also need the nature of the particle at the time of its decay, typically
given by one or more of the decay particles.  In the case of lifetime 
comparisons, one may use untagged mesons, and gain the increase in efficiency 
implied.

Universally, tagging of $D^o$'s is done by examining only those $D^o$'s
which are the decay products of charged $D^{*'}$s.   In this case, the
strong decay of the $D^{*\pm} \rightarrow D^o \pi^{\pm}$ gives the nature of
the charm quark in the $D^o$, since it is the same as that in the $D^{*}$,
and is marked by the $D^{*}$ charge and that of the daughter charged
pion.  Clearly, using only such $D^o$'s reduces the size of the $D^o$-sample
available for study.  Fortunately, the production of $D^{*'}$s is frequent in 
charm events, and the $ D^o \pi^{\pm}$ decay is both copious and easy to observe.

To date, the observed decays used in mixing studies are:  \\
Hadronic Decays:
$D^o_{tag} \rightarrow K \pi $ and $ \rightarrow K\pi\pi\pi $
\\
Semileptonic Decays:
$D^o_{tag} \rightarrow K\mu\nu $ and $ \rightarrow K e \nu $
\\
Lifetime Differences:
$D^o \rightarrow K^+K^-$, $K_s^o \phi$, and $ K\pi $.  Comparison
of decay rates can be made between the CP eigenstates, or to the mixed 
state $ D^o \rightarrow K\pi $.

\subsection{Hadronic Decays of Tagged $D^o$ Mesons}

In hadronic decays, it is possible to reach the final state which would
come from mixing by doubly-Cabibbo-suppressed decay.  Such
doubly-Cabibbo-suppressed decays are expected at about the level of
today's limits on mixing.  Thus, the analyses must take these decays into
account.  The methods used involve a maximum-likelihood fit to a sample
of events which have the characteristic charge correlations for mixing.
The fit function for the signal includes the signature for the 
$D^o$ decay (a Gaussian-function distribution for the effective mass of the 
$D^o$ decay products, $G_D(M)$), the signature for the tagging $D^*$ decay 
(a Gaussian-function distribution in the mass difference in the $D^*$ decay, 
$G_{D*-D}(Q)$), and the separate proper-time distributions for probabilities 
coming from mixing, from the doubly-Cabibbo-suppressed mechanism, and from the
interference of mixing and doubly-Cabibbo-suppressed amplitudes.  
The proper time of decays is needed to separate origins in mixing from 
double-Cabibbo-suppression.  The backgrounds, $B(M,Q,t)$, are also 
parameterized in terms of the same variables as used for the signal.
Expressions for the terms in the  maximum-likelihood function are given 
in Eqns. \ref{maxlike} to \ref{where4}.
\begin{equation}
N(M,Q,t) = G_D(M) * G_{D*-D}(Q) * \epsilon(t) S(t)
+ B(M,Q,t)    
\label{maxlike}
\end{equation}
Where, for the signal part:
\begin{equation}
S(t) = [N_{MIX}*f_{MIX}(t) + N_{DCSD}*f_{DCSD}(t) + N_{INT}*f_{INT}(t)]
\label{where1}
\end{equation}
\begin{equation}
f_{MIX}(t)  =  t^2 * e^{-\Gamma t} 
\label{where2}
\end{equation}
\begin{equation}
f_{DCSD}(t) =    e^{-\Gamma t}     
\label{where3}
\end{equation}
\begin{equation}
f_{INT}(t)    =  t * e^{-\Gamma t} 
\label{where4}
\end{equation}
and the detection efficiency, $\epsilon(t)$, may be a function of the proper
time.

We are now entering the time when the interference term may provide the
greatest sensitivity to mixing, since the square of the limit on the 
mixing amplitude is now smaller than the visible doubly-Cabibbo-suppressed
rate.  Of course, such sensitivity depends on the phase between the
Cabibbo-favored and doubly-Cabibbo-suppressed amplitudes.  Yesterday's 
background may be tomorrow's signal enhancer!

\begin{figure}[t]
\vspace{9.0cm}
\includegraphics{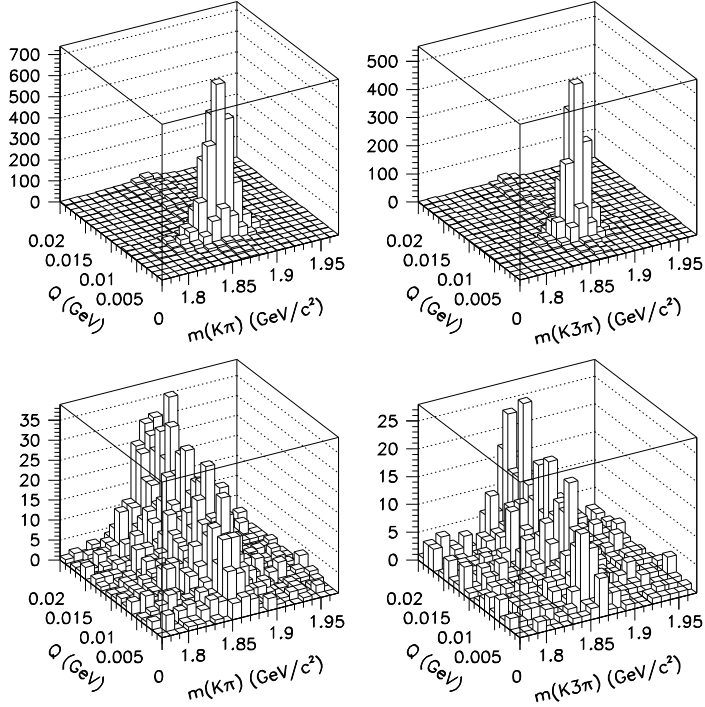}
\caption{\it  The E791 signals used to establish a limit on charm mixing.
Both $D^o \rightarrow K\pi$ (left) and $D^o \rightarrow K\pi\pi\pi$ (right)
are shown.  The Cabibbo-favored signals are shown in the top figures, the
opposite sign correlations on the bottom.}
\label{e791-fig}
\end{figure}

\begin{figure}[t]
\vspace{9.0cm}
\includegraphics{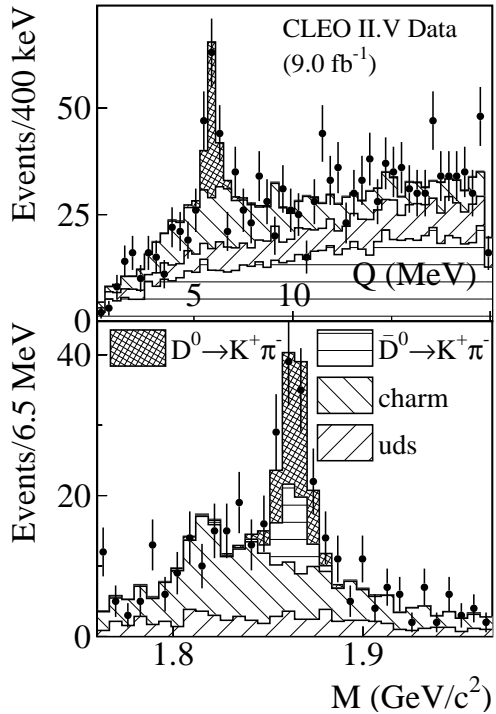}
\caption{\it  The CLEO wrong-sign signal used to establish their 
$D^o \rightarrow K\pi$ limit on mixing.  In the top plot, 
$M$ is within $14 MeV$ of the nominal $D^o$ mass, and for the bottom plot 
$Q$ is within $500 KeV$ of the nominal value.
The signal and various backgrounds are indicated in the figure by hatching, 
with the data given by the points with error bars.}
\label{cleo-fig}
\end{figure}

The recent mixing results are shown in Table \ref{extab1}.  The earliest
of these comes from the full data set of Fermilab's E791 charm hadroproduction 
experiment.\cite{e791-mx1} These results are 
final, and published.  Distributions are shown for the E791 hadronic-decay 
study\cite{e791-mx2} in Fig. \ref{e791-fig}.
The figure gives an indication of the kind of distributions which enter
the maximum-likelihood fits using Eqns. \ref{maxlike} to \ref{where4}.
The ALEPH data comes from the full $Z^o$ data from 1991-1995 running at
LEP, and have also been published.\cite{ALEPH}  The CLEO II.V preliminary 
result,\cite{cleo-mx1}\cite{cleo-mx2}\cite{cleo-mx3} which comes from the data 
shown in Fig. \ref{cleo-fig}, is also from their full data set of $ 9 fb^{-1}$.  
The first results from Fermilab's photoproduction experiment, FOCUS/E831, 
are expected soon,\cite{focus-mx1}

The CLEO result is the most constraining, at the level of 0.05 $\%$,
coming from the fact that the wrong-sign events (those characteristic of
mixing and of doubly-Cabibbo-suppressed decays) appear at short proper
decay-time.  The short lifetime of these events strongly rejects large
constructive interference between mixing and DCSD.  As seen in 
Table \ref{extab1}, some earlier mixing analyses assumed no CP violation; 
and there are results quoted with the interference term arbitrarily 
set to zero.  The more general fits, allowing the most general solution, 
typically result in looser quoted constraints on mixing.  The excellent
CLEO acceptance at short proper-lifetime relative to that at fixed-target 
experiments also makes the CLEO result less sensitive to the generality
of the fit used.

\subsection{Semileptonic Decays of $D$ Mesons}

Semileptonic decays have the advantage that there is no 
doubly-Cabibbo-suppressed decay to obscure a mixing interpretation.
Tagging of the initial state is still required, of course.  While the E791
results\cite{e791-mx1} are available and listed in Table \ref{extab1}, only 
the promise of the FOCUS/E831 data set is known.\cite{focus-mx1}  They project 
a 90\% CL upper limit of 0.1\% after combining their electron and muon mode
data, and assuming that "they observe precisely zero background-subtracted
events in their wrong sign signal region."\cite{focus-mx1}   We anxiously await
the result of their full data set.  

\subsection{Lifetime Differences Among Various CP Mixtures of Neutral $D$
Mesons}

Mixing can appear if there is either a difference in the masses of the
CP eigenstates $\Delta m$ or if there is a difference in the decay rates 
$\Delta\Gamma$ (Eqn. \ref{rmix}).
\begin{equation}
r_{MIX} = \frac{(\Delta m)^2}{2\Gamma^2}+\frac{(\Delta\Gamma)^2}{8\Gamma^2}
= \frac{1}{2}(x^2 + y^2),
\label{rmix}
\end{equation}  
\begin{equation}
\Gamma = (\Gamma_1 + \Gamma_2)/2 
\label{gamma}
\end{equation}  
where $\Gamma_1$ is for CP-even states,
$\Gamma_2$ for CP-odd states, and
\begin{equation}
\Delta\Gamma = \Gamma_1 - \Gamma_2.
\label{deltagamma1}
\end{equation}
$\Gamma_1$ applies to $D^o \rightarrow K^+K^-$ and $\pi^+\pi^-$ and
$\Gamma_2$ applies to  $D^o \rightarrow K_s^o \phi,  K_s^o \omega$, and  
$K_s^o \rho$.
$\Gamma$ applies to  $D^o \rightarrow K \pi$, if CP is conserved.
And, then  \\
\begin{equation}
\Delta\Gamma  =   2 (\Gamma_{KK}-\Gamma_{K\pi})         
               =   2(\Gamma_{K\pi}-\Gamma_{K\phi})  
               =    \Gamma_{KK}-\Gamma_{KVector}          
\label{deltagamma2}
\end{equation}
\\
The E791 measurement\cite{e791-mx3} gives
\begin{equation}
\Delta\Gamma = 2(\Gamma_{KK} - \Gamma_{K\pi}) = (0.04\pm0.14\pm0.05)ps^{-1}
\label{deltagamma3}
\end{equation}
This directly measured $\Delta\Gamma$ limit is more constraining than
that which is obtained from Eqn. \ref{rmix}, the indirect limit
from no mixing assuming $\Delta m$ is zero.  Results including the CP-odd
decays are anticipated from CLEO and FOCUS.

\begin{table}[t]
\centering
\caption{ \it Recent Results on Charm Mixing. Values for $r_{MIX}$, x, x', y, and
y' are given in \%;
values for $\Delta\Gamma$ , in $ ps^{-1}$.  $D^o_{tag}$ refers to $D^o$'s whose 
particle-antiparticle nature at birth is known (tagged).  Confidence levels as are 
indicated explicitly when not at 90\%. The CLEO result has been updated from what was
presented at the workshop.\cite{cleo-mx3}}
\vskip 0.1 in
\begin{tabular}{|l|c|c|c|} 
\hline
    Decay Mode                    & Results (\%, $ps^{-1}$)       &
$ 90 \%$ CL Limit
& Exp. \\
\hline
\hline
$D^o_{tag} \rightarrow K\pi      $&                                
&$r_{MIX}<0.92  $
&ALEPH     \\
                                  &                                & ($95\%$ CL)
&No CP V.   \\
                                  &                                &
&No Interf.\\
$D^o_{tag}\rightarrow  K\pi\pi\pi$& $r_{MIX}=0.21 \pm0.09 \pm0.02$ & 
 $r_{MIX}<0.36$ 
&E791      \\
                                   &                               &        
&No CP V.   \\
                                   &                               &
&No Interf. \\   
\hline
$D^o_{tag} \rightarrow K\pi$       &$x' = 0.0 \pm 1.5 \pm 0.2  $   &
$r_{MIX}<0.05$
&CLEO II.V  \\
                                   &$y' =-2.5_{-1.4}^{+1.5}\pm 0.3$& ($95\%$ CL) 
&           \\
\hline
$D^o_{tag}\rightarrow K\pi$        &$r_{MIX}=0.18_{-0.39}^{+0.43}\pm0.17$&
$r_{MIX}<0.94$
&E791       \\
and                                &                                  &         
&$\overline{D}^o \rightarrow D^o$ \\ 
$D^o_{tag} \rightarrow K\pi\pi\pi$ &$r_{MIX}=0.70_{-0.53}^{+0.58}\pm0.18$&
$r_{MIX}<1.31$
& E791  \\
                                   &                                  &      
&$D^o \rightarrow \overline{D}^o$ \\
\hline
$D^o_{tag}\rightarrow K\mu\nu  $   &$r_{MIX}= 0.06_{-0.40}^{+0.44} $  & 
$r_{MIX}<0.50$
& E791  \\
$D^o_{tag}\rightarrow K e \nu  $   &$r_{MIX} = 0.16_{-0.37}^{+0.42} $ &    
&       \\
\hline
$D^o \rightarrow K K  $            &$\Delta\Gamma =0.04 \pm 0.14\pm 0.05 $
&$-0.20<\Delta\Gamma$  & E791\\
                                 &                                  &
$\Delta\Gamma < 0.28 $           & No CP V. \\
                         & $y = 0.8_{-1.0}^{+2.9} $       & $-2.4<y<5.6$
&       \\
\hline
\end{tabular}
\label{extab1}
\end{table}

\subsection{What Models Are Tested in Charm Mixing?}

As we have noted, typical Standard-Model predictions are many 
orders of magnitude smaller than the results in Table 1.  However, there
are many non-standard models which predict charm mixing at, or even
above, the current limits.  These models include those with light 
leptoquarks, SUSY particles, fourth-generation quarks, and Higgs particles.
In each case, these objects occur in internal loops and their effects are
virtual, if observable.  And, in spite of calling such virtual particles 
light, their masses are still much above the mass reach of
direct-production experiments, even at today's highest energy machines.
Harry Nelson has compiled over thirty Standard-Model and
non-Standard-Model predictions,\cite{nelsonpred} and promised to keep his
compilation updated.  What he shows is the largest mixing rate for each
model assuming "standard" couplings.  In fact, a more detailed summary
cannot be presented in a single parameter such as the rate, since each
prediction depends not only on the mass of the virtual particle involved,
but also on its couplings to the charm and other quarks of the final
state.  Examples of two-dimensional exclusion regions are
given for representative models by Gustavo Burdman\cite{burdman} and
Joanne Hewett.\cite{hewett}  What we see are limits on otherwise allowed
parameters, but more or less at the extremes of what we might otherwise
expect.  That is, charm measurements do limit the parameter space of
allowed particles beyond the Standard Model.  However, we are just getting
into the most interesting regions now.  The future could be much more
exciting.

\section{CP Violation}

There are four types of searches for CP violation: three for asymmetries 
in the decay rates of charm particles and antiparticles and one for 
differences in density distributions in Dalitz plots for decaying
particles and antiparticles.  The decay-rate asymmetries may be due to:
(1) particle-antiparticle mixing, (2) direct CP violation in particle and 
antiparticle decays to identical final states, and (3) direct CP violation 
in decays to different final states (i.e., opposite charges).  The first 
two of these occur only for neutral meson decays.  The second is only 
possible for Cabibbo-suppressed decays.  The third is pursued in
charged-meson decay.  

The ideal situation for observing CP violation occurs when there are two
routes to a given final state, the amplitudes describing the routes have a
significant relative phase, and there is a significant difference in 
the strong phases of the final-states depending on the route.  In addition,
it is best if the amplitudes for the two routes have comparable magnitude.
We can see these features if we write the generic, total amplitude for 
decay via two mechanisms as
\begin{equation}
A = A_1  e^{i \delta_1} + A_2  e^{ i \delta_2 }
\label{amp}
\end{equation}
where the $ A_i$ are the (complex) weak-decay amplitudes and the $\delta_i$
are the relevant strong-interaction phases.  The CP conjugate amplitude is
\begin{equation}
A = A_1^* e^{i \delta_1} + A_2^* e^{ i \delta_2 }
\label{amp-conj}
\end{equation}
Then, the CP violation is observed as an non-zero asymmetry calculated from 
the decay rates of the particle and antiparticle:
\begin{equation}
A = \frac{2\it{Im}A_1 A_2^*sin(\delta_1-\delta_2)}
{ |A_1|^2+|A_2|^2+2\it{Re}A_1 A_2^* cos(\delta_1-\delta_2)}
\label{A}
\end{equation}
Ideally, i.e., for large measurable asymmetries, one would like $|A_1|$ and
$ |A_2|$ to be comparable in size, and both the phases of the weak
amplitudes $A_i$ and of the strong phases $\delta_i$ should be quite
different. 

\subsection{Rate Asymmetries}

For neutral $D$-mesons, CP violation may occur via particle-antiparticle
mixing and via direct CP violation.  In mixing, the two amplitudes involved
are those relating to the particle and antiparticle decays to the final
state.  For direct CP violation, the two amplitudes come from different
mechanisms for the meson to decay directly to the given final state.  Two
such amplitudes are those for the spectator and penguin mechanisms.  

Charged $D$-mesons can have only
direct CP violation.  As an example of direct CP violation, consider the
decay $D^+ \rightarrow K^*(892) K$.  In this case, the spectator amplitude
involves the product of CKM matrix elements $V_{cs}^*V_{us}$, while the
penguin amplitude involves $ V_{cb}^*V_{ub} $.  Thus, there are two weak
amplitudes with a phase difference given by the CKM matrix phases.  In
addition, the spectator process involves both isospin 1/2 and 3/2 
amplitudes.  The penguin process is pure isospin 1/2.  The strong phases of
these isospin amplitudes can have very different values due to 
final state interactions in kinematic regions with 
nearby resonances.  In fact, Alain  LeYaouanc has predicted 
$ A_{CP}$ to be $\sim 10^{-3}$\cite{LeY} and Franco Buccella has predicted 
$\sim (1.4-2.8) x 10^{-3} $.\cite{buccella}  In general, final-state 
interactions (rescattering effects) are important for charm.  For example, 
\begin{equation}
B(D^o \rightarrow K^o K^o)/ B(D^o \rightarrow K^+ K^- ) =  0.24 \pm 0.09 
\label{Bratio}
\end{equation}
where a ratio more nearly unity is expected if only phase-space differences
are considered.

\begin{figure}[t]
\vspace{9.0cm}
\includegraphics{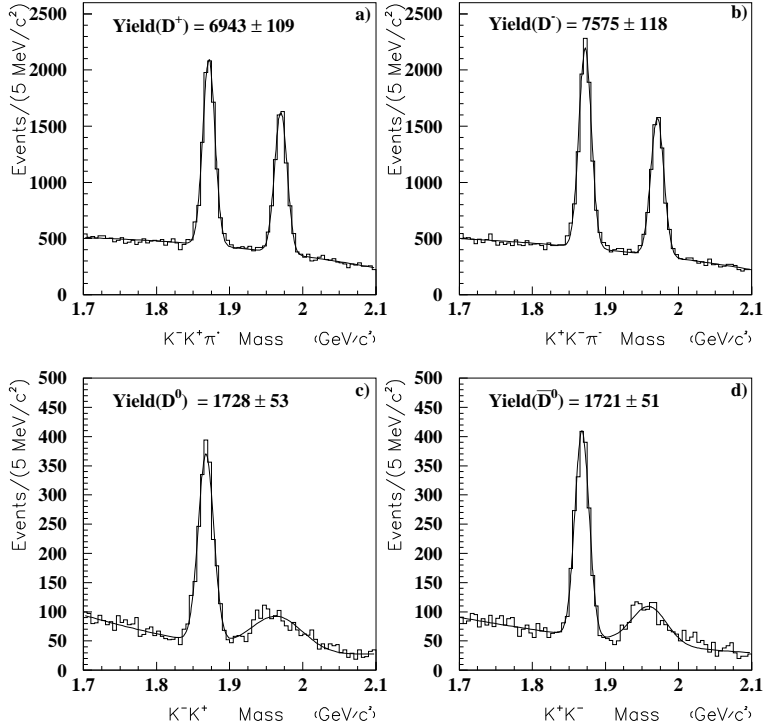}
\caption{\it
Invariant mass, from the total FOCUS data sample, for 
a) $D^+\rightarrow K^- K^+ \pi^+$, b) $D^-\rightarrow K^- K^+ \pi^-$,
c) $D^o\rightarrow K^- K^+$, and $\overline{D}^o\rightarrow K^- K^+$.}  
\label{focus-cp-1}
\end{figure}

As an example of the experimental method, consider the effective mass plots 
from FOCUS for the Cabibbo-suppressed decays of the charged and neutral $D$
mesons shown in Fig. \ref{focus-cp-1}.  The peak on the left of each
figure is due to the relevant $D$ decay.  There is an immediate observation 
that the numbers of mesons and antimesons are unequal in each case.  However,
one must first take account of the differences in production rates.  This
is done by taking the asymmetry of ratios; i.e., of each signal normalized 
to its observed Cabibbo-favored decay.  Table \ref{extab2} lists the recent 
results on CP violation searches.  The results come, again, from
E791\cite{e791-cp1}\cite{e791-cp2} and from FOCUS.\cite{focus-cp}

\subsection{Differences in Dalitz Plots of Particle and Antiparticle Decays}

\begin{figure}[t]
\vspace{9.0cm}
\includegraphics{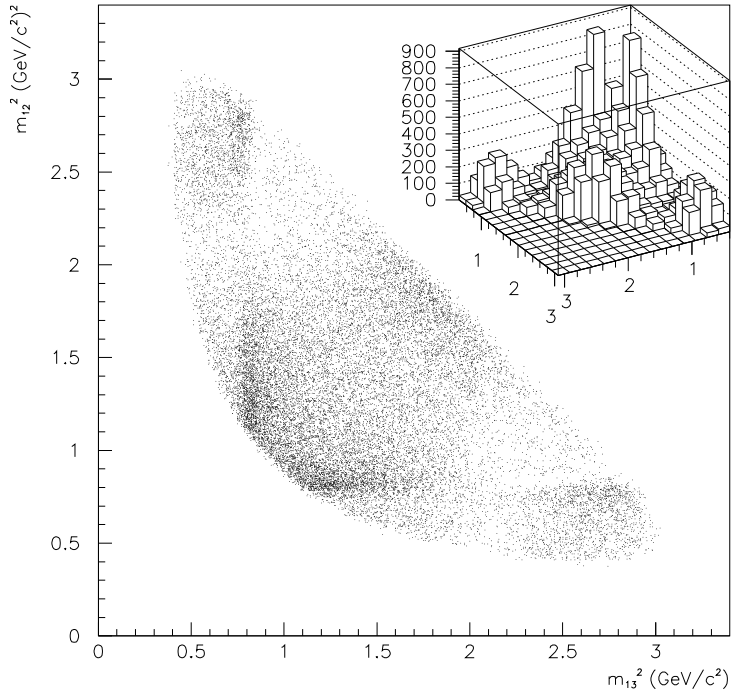}
\caption{\it  Dalitz-plot distribution of events for 
$D^+\rightarrow K^- \pi^+ \pi^+$ from Experiment E791.}  
\label{Dalitz}
\end{figure}

In the literature, there are no results quoted so far for differences in Dalitz
plots as a search for CP violation.  In general, experimenters use such comparisons
to look for instrumental asymmetries which must be found and removed -- if they 
look at all.  In the case of charm, typically, when there is more than a single
Standard Model contribution to a decay channel, there are no phase differences
expected in the amplitudes for particle and antiparticle.  In order to be seen, 
any new-physics contribution
should contribute to one of the possible amplitudes so that there is a
net phase difference available for the interference term in the decay rate.
It is instructive to look at the Dalitz plot (Fig. \ref{Dalitz}) for E791 data
on the decay 
$D^+ \rightarrow K^- \pi^+ \pi^+$.  This  Dalitz plot shows very clearly 
what interference with a coherent phase difference can do in a Dalitz plot.  
There is a large  $K^*$ contribution and a much broader contribution evident
in the plot.  Note the change from constructive to destructive interference 
as one moves from one side of the $K^*$ mass squared to the other.  If there 
were a difference in this pattern between $D^+$ and  $D^-$ decays, we would have
evidence of CP violation.  In fact, the place to look would be in Cabibbo-suppressed
modes where any CP-violation signal is more likely.  Although the available
statistical precision of the data does not allow such visual clarity as that
in Fig. \ref{Dalitz}, we may hope to achieve this level with future charm data.

\subsection{What Models Are Tested in Searches for CP Violation?}

The typical $90\%$ confidence level limits shown in Table \ref{extab2} are at 
the $10^{-1}$ level.  As noted, the Standard-Model asymmetry predictions 
from higher order and long range processes are at the $10^{-3}$ level.  Thus, 
there is a so-called "window of opportunity" of two orders of magnitude in which 
non-Standard-Model effects might be observed.  Such effects could be due to 
processes in models with SUSY particles, left-right symmetric particles, or 
extra Higgs particles.\cite{Gounder:1994cd}

\begin{table}[t]
\centering
\caption{ \it Recent Results on CP Violation.   $D^o_{tag}$ refers to $D^o$'s
whose particle-antiparticle nature at birth is known (tagged).  The FOCUS 
results are preliminary, and their reported errors are just the statistical errors.}
\vskip 0.1 in
\begin{tabular}{|l|c|c|c|} \hline
              Decay Mode  & Result (\%)  &$90 \% CL Limit (\%)$  &
Experiment \\
\hline
\hline
$D^+ \rightarrow K K \pi$              &$-1.2 \pm 1.1$ &
&FOCUS
\\
                                       &$-1.4 \pm 2.9$ &$-6.2<A_{CP}<3.4$
& E791\\
\hline     
$D^+ \rightarrow \phi \pi $          &$-2.8 \pm 3.6$ &$-8.7<A_{CP}< 3.1$
&E791\\
\hline     
$D^+ \rightarrow K^*(892) K$         &$-1.0 \pm 5.0$ &$-9.2<A_{CP}< 7.2$&
E791\\
\hline     
$D^+ \rightarrow  \pi \pi \pi $      &$-1.7 \pm 4.2$ &$-8.6<A_{CP}< 5.2$&
E791\\
\hline     
$D^o_{tag}  \rightarrow  K K $       &$ 0.0 \pm 2.2$ &                       
&FOCUS 
\\
                                  &$-1.0 \pm 4.9 \pm 1.2$&$- 9.3< A_{CP}<7.3$
&E791\\ 
\hline     
$D^o_{tag} \rightarrow \pi \pi$      &$-4.9 \pm 7.8 \pm 3.0$&$-18.6<A_{CP}<8.8$
&E791\\
\hline     
$D^o_{tag} \rightarrow K\pi\pi\pi$   &$ 1.8 \pm 2.3 \pm 0.2$&$-5.5<A_{CP}<1.9 $
&E791\\
\hline
\end{tabular}
\label{extab2}
\end{table}

\section{Overview of What's Been Achieved}

\begin{table}[t]
\centering
\caption{\it Numbers of Observed Events by Analysis.  CLEO results in 
parentheses are from only $5.6 fb^{-1}$ of $9.1 fb^{-1}$ collected.  
The ALEPH collaboration has recently reported a result based on
$1,039\pm33 D^o_{tag} \rightarrow K\pi $ events, not listed in the table to
save space.   The CLEO CP-violation result has been added since the 
workshop.\cite{cleo-mx3}} 
\vskip 0.1 in
\begin{tabular}{|l|c|c|c|c|} 
\hline
Physics  & Decay Mode        & E791 & CLEO II.V & FOCUS \\
 Topic   &                   &      &           &       \\
\hline
\hline
Mixing   &$D^o_{tag} \rightarrow K\pi    $&$  5,643\pm77$& $16,126\pm127$ 
&              \\
         &$D^o_{tag}\rightarrow K\pi\pi\pi$&$  3,469\pm60$&           
&              \\
\hline
Mixing   &$D^o_{tag} \rightarrow K\mu\nu $&$  1,267\pm44 $&   
&$ 16,522\pm200$\\
         &$D^o_{tag} \rightarrow Ke\nu   $&$  1,237\pm45 $&              
&$  (92 \%)    $\\   
\hline
$\Delta\Gamma$&$D^o \rightarrow KK        $&$  3,200\pm57 $&$(1,300\pm40) $
&               \\
             &$D^o \rightarrow K\pi      $&$ 35,400\pm206$&$(19,000\pm140)$
&$ 99,800\pm340$\\   
             &$D^o \rightarrow K_s^o\phi $&               &$(3,000\pm60)$
&              \\
\hline
CP Vio.     &$D^+ \rightarrow KK\pi     $&$ 2,296\pm65  $&
&$14,518\pm161$\\
             &$D^+ \rightarrow \phi\pi   $&$ 1,072\pm38  $&
&              \\
             &$D^+ \rightarrow K^*K^+    $&$  530\pm26  $&
&              \\
             &$D^+ \rightarrow \pi\pi\pi $&$  1,548\pm64 $&
&              \\
             &$D^+ \rightarrow K\pi\pi   $&$51,479\pm272 $&
&$ 146,497\pm426$\\
\hline
CP Vio.     &$D^o_{tag} \rightarrow K K $&$   609\pm29  $&  $ 3,023 \pm 66$
&$ 3,449\pm 74 $\\
             &$D^o_{tag} \rightarrow \pi\pi$&$ 343\pm25  $&
&               \\
             &$D^o_{tag} \rightarrow K\pi\pi\pi$&$ 3,409\pm62 $&
&               \\
             &$D^o_{tag} \rightarrow K\pi$&$13,273\pm129 $& $13,527 \pm 116$
&$ 39,206\pm211$\\
\hline
\end{tabular}
\label{extab3}
\end{table}

In order to understand the increased sensitivity achieved so far, we
need to look at the numbers of observed events in each of a variety of 
physics analyses.  Table \ref{extab3} gives the numbers for the latest round
of experiments on mixing and CP violation.  Since some of the data sets are
not fully analyzed, we should extrapolate each experiment's numbers to the
size of its full recorded set.  At the same time, it is best to make
comparisons in an equivalent way, independent of the varied background level
present in each experiment.  We do this in each case by taking the square of 
the ratio of the number of events divided by the quoted statistical error in 
that number.  Such figures-of-merit are presented in Table \ref{extab4}.

From the numbers in Table \ref{extab4}, it appears that CLEO and FOCUS will 
have the best results from existing data sets for most topics.  Between the two
experiments, the results will improve over hadroproduction experiment E791 by 
factors of three for hadronic modes and ten for semileptonic modes.  We may expect 
the errors to scale as the inverse of the square root of the numbers of events.  
Systematic errors will need to be reduced accordingly, a task which is easier 
at $e^+e^-$ machines where the signals often appear with less background.  
When the systematic uncertainties can be controlled with the increased amount 
of data, the physics reach will improve by factors of the square root of three 
to the square root of ten.  For future data sets, physics reach will also 
scale like the square root of these reduced numbers of reconstructed decays.

\begin{table}[t]
\centering
\caption{ \it Numbers of Equivalent Pure Decays Observed by Analysis (scaled
to full data sets where needed).  The CLEO CP-violation result has been added
since the workshop.\cite{cleo-mx3}}
\vskip 0.1 in
\begin{tabular}{|l|c|c|c|c|c|} \hline
Physics & Decay Mode  & ALEPH & E791 & CLEO II.V & FOCUS \\
 Topic  &             &       &      &           &       \\     
\hline
\hline
Mixing       &$D^o_{tag} \rightarrow K\pi     $&$1000$&$5,400$&$16,000$& \\     
             &$D^o_{tag} \rightarrow K\pi\pi\pi$&  &$3,300 $&       &
\\     
\hline
Mixing       &$D^o_{tag} \rightarrow K\mu\nu $&    &$  750 $&       &
$7,400$\\   
             &$D^o_{tag} \rightarrow Ke\nu   $&    &$  760 $&       &
\\   
\hline
$\Delta\Gamma$&$D^o \rightarrow KK           $&    &$3,150 $&$ 1,700$&
        \\  
             &$D^o \rightarrow K\pi          $&    &$29,500$&$30,000$&
$86,000$\\   
             &$D^o \rightarrow K_s^o\phi     $&    &$      $&$ 4,100$&
\\  
\hline
CP Vio.      &$D^+ \rightarrow KK\pi         $&    &$1,250 $&        &
$ 8,100$ \\
             &$D^+ \rightarrow \phi\pi       $&    &$  800 $&        &       \\  
             &$D^+ \rightarrow K^*(892)K^+   $&    &$  420 $&        &       \\      
             &$D^+ \rightarrow \pi\pi\pi     $&    &$  590 $&        &       \\    
             &$D^+ \rightarrow K\pi\pi       $&    &$36,000$&        &
$120,000$\\  
\hline
CP Vio.     &$D^o_{tag} \rightarrow K K      $&    &$  440 $& $2,100$&
$ 2,200$ \\  
             &$D^o_{tag} \rightarrow \pi\pi    $&  &$  190 $&        &       \\  
             &$D^o_{tag} \rightarrow K\pi\pi\pi$&  &$3,000 $&        &       \\   
             &$D^o_{tag} \rightarrow K\pi      $&  &$10,500$&$13,500$&
$35,000$ \\  
\hline
\end{tabular}
\label{extab4}
\end{table}

\section{Expectations for the Future}
    
We have seen excellent signal to (well understood) backgrounds in today's
charm decay experiments.  This has led to real improvements in sensitivity 
to new physics.  The progress has been the result of precision
reconstruction of production and decay vertices, excellent kinematic
resolution, and increasingly large data samples.  Some of this has come
from dedicated charm experiments; other progress is the byproduct of
B-motivated experiments.  Since we may have seen the last of dedicated charm
experiments [Can we hope still for a t-charm Factory?], we need to understand
what may be expected from future B-motivated experiments.  

There is the potential for $10^7$ reconstructed
charm decays from  B factories (and COMPASS); also the potential
for $10^{8-9}$ from BTeV (and LHC-b?).  Even though hadron environments
may be harder, the production rate, coupled to capable detectors, can win in
the end.  This has been shown by E791.  However, triggers will have to
allow/encourage charm data to be taken! As it is, charm events may be the
worst enemy of B-experiment triggers.  Often, B experiments actively try
to minimize the charm events recorded.

\section{Summary and Conclusions}

Charm experiments have reached the level of $10^6$ reconstructed meson
decays.  FOCUS holds the record in this regard today.  So far, there is 
no evidence for either mixing or CP violation in the charm sector.  

The march toward increasing numbers of well-reconstructed decays with 
well-understood backgrounds has led to decades of increased sensitivity
over the last years.  There is hope for continued progress in this direction.
However, this hope depends mostly on results coming as a side benefit from
the major B efforts coming on line, especially those whose on-line event
selection allows charm data to be taken.

The mass reach for new physics sources via virtual processes in charm
decay greatly exceeds what can be directly produced now, or in the
foreseeable future.  Who knows, new physics could be just around a charmed
corner.

\section{Acknowledgements}

I want to begin by acknowledging my colleagues on E791 who have taught me so
much of what I know about the subjects reported here.  In addition, for
help with the data presented in this review, I have very much benefited from
the help of David Asner (CLEO II.V), Carla Gobel, Jean Slaughter, Mike Sokoloff, 
and Ray Stefanski (E791), and Stefano Bianco, Daniele Pedrini, and Jim Wiss (FOCUS).
I also thank the organizers of this workshop whose efforts have shown both in
the workshop's physics breadth and intensity, and how truly smoothly 
everything has gone.

\end{document}